%% file: the_paper.tex
\documentclass[12pt]{article}
\usepackage{pic02}
\usepackage{epsfig}
\usepackage{colordvi}
\usepackage{hyperref}
\usepackage{url}
\usepackage{graphicx}

\newcommand{\Figures}{Figures/}
\newcommand{\Journal}[4]{{#1} {\bf #2}, #3 (#4)}

\newcommand{\NPB}{{\em Nucl. Phys.} B}
\newcommand{\PLB}{{\em Phys. Lett.}  B}
\newcommand{\PRL}{\em Phys. Rev. Lett.}
\newcommand{\PRD}{{\em Phys. Rev.} D}

\newcommand{\PTP}{{\em Prog. Theor. Phys.}}

\newcommand{\be}{\begin{equation}}
\newcommand{\ee}{\end{equation}}
\newcommand{\bea}{\begin{eqnarray}}
\newcommand{\eea}{\end{eqnarray}}

\newcommand{\Ampl}{{\cal A}}
\let \bar=\overline
\let \to=\rightarrow
\let\mb=\boldmath

\input mytex.tex

\input babar.tex

\begin{document}

\title{\bf SOME DEVELOPMENTS IN LIGHT QUARK SPECTROSCOPY
}
\author{
Brian T. Meadows \\
{\em Physics Department, University of Cincinnati, Cincinnati,
 OH 45221-0011}}
\maketitle

%
%
%
%
%
%
\vspace{3.5cm}
%

\baselineskip=14.5pt
\begin{abstract}
\input{abstract.tex}
\end{abstract}
\newpage

\baselineskip=17pt

\section{Introduction}
\label{sec:intro}
\input{intro.tex}

\section{Data from Charmed Meson Decays}
\label{sec:ddk}
\input{ddk.tex}

\subsection{The Isobar Model}
\label{sec:isobar}
\input{isobar.tex}

\section{A Hint of a ``$\kappa$" Meson.}
\label{sec:kappa}
\input{kappa.tex}

\section{Evidence for a Low Mass ``$\sigma$".}
\label{sec:sigma}
\input{sigma.tex}
\section{New $f_{\z}(980)$ Data.}
\label{sec:f0}
\input{f0.tex}
\section{Other $f_{\z}$ Results}
\label{sec:f0_other}
\input{f0_other.tex}

\section{Summary}
\label{sec:summary}
\input{summary.tex}

\section{Acknowledgements}
This work was supported by NSF award number:0203262.
The authors gratefully acknowledge valuable discussions with
my E791 colleagues and with W. Dunwoodie.

\end{document}

%% file: mytex.tex
\newcommand{\ssc}{\scriptscriptstyle}

\newcommand{\etc}{{\sl etc~}}

\newcommand{\z}{_{\circ}}

%% file: babar.tex
\let\mathrm=\rm

\newcommand{\Kz}{K^{\circ}}

\newcommand{\etal}{{\sl etal}~}

\newcommand{\fz}{f_{0}}
\newcommand{\az}{a_{0}}

\newcommand{\piz}{\pi^{\circ}}
\newcommand{\pip}{\pi^+}
\newcommand{\pim}{\pi^-}
\newcommand{\Kp}{K^+}
\newcommand{\Km}{K^-}

%% file: abstract.tex
Among the many unresolved questions in light quark spectroscopy,
the underlying structure of the scalar mesons and the
identification of states with a gluonic content rank high.
Recently, new information has come from $\phi$ radiative decays,
$J/\psi$, $\tau$, $D$ and $D_s$ meson decays.  Other papers in this
conference review radiative transitions of $\phi$ and $J/\psi$.
This paper discusses new information on the scalar sector
primarily that from decays of $D$ and $D_s$ mesons.

%% file: intro.tex
One important reason for continuing to study light quark spectroscopy is
to search for glueballs and exotics predicted by QCD.
\\[-24pt]
\newlength{\fh}
\setlength{\fh}{0.35\textwidth}
\begin{figure}[ht]
\begin{minipage}[t]{0.38\textwidth}
 \vskip20pt\hskip0.25\textwidth\bf (a)\vskip-20pt
 \centerline{%
 \epsfig{file=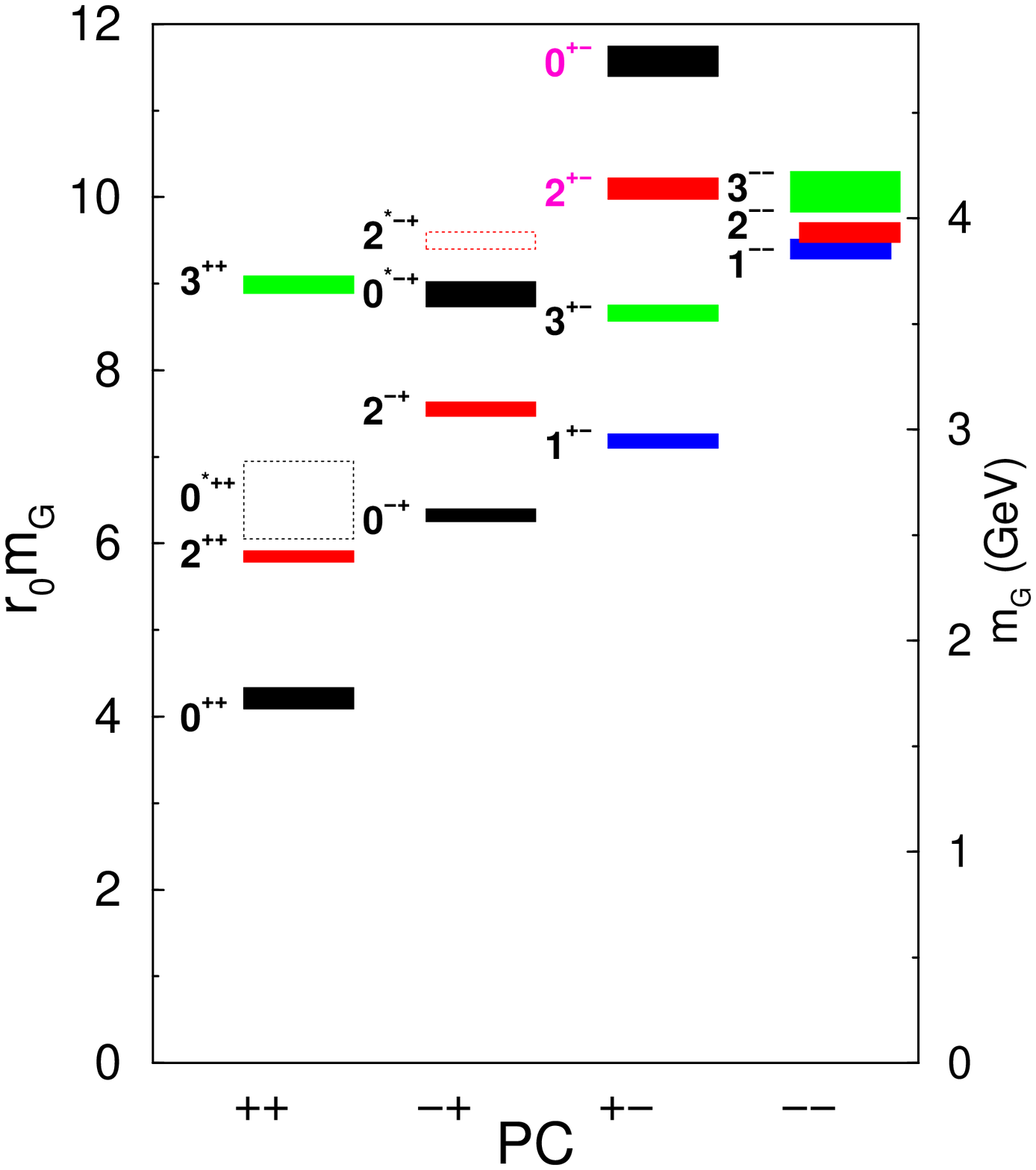,
      height=\fh,
      width=0.97\textwidth,
      angle=0}}
  \par\vspace{0pt}
\end{minipage}
\hspace{0.08\textwidth}
\begin{minipage}[t]{0.51\textwidth}
 \vskip20pt\hskip0.85\textwidth\bf (b)\vskip-20pt
 \centerline{%
 \epsfig{file=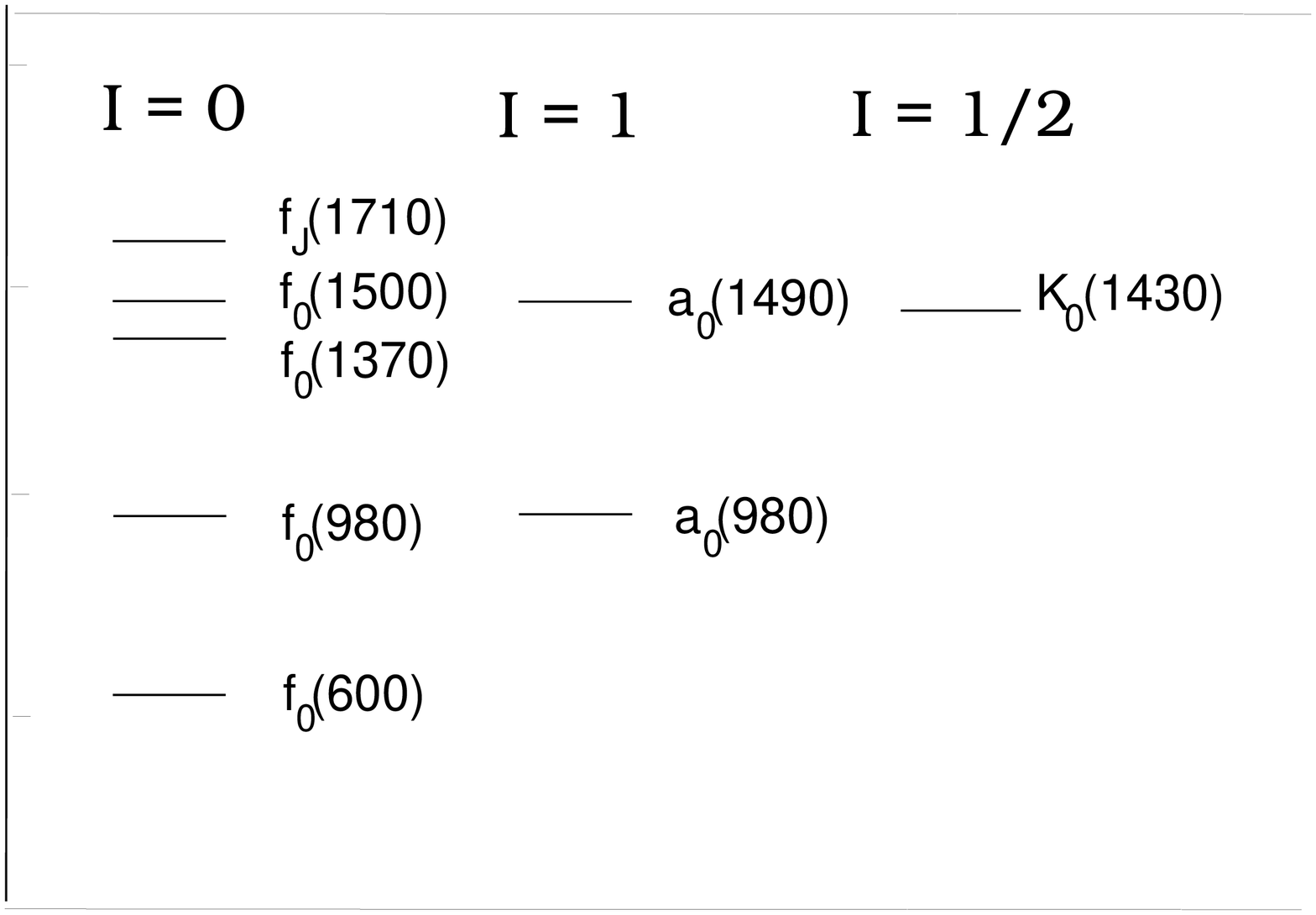,
      height=\fh,
      width=0.97\textwidth,
      angle=0}}
 \par\vspace{0pt}
\end{minipage}
\caption{\sl (a) Spectrum predicted from unquenched QCD for glueball
             states from a recent summary by Morningstar
             \cite{morningstar}.
             (b)
             Scalar states in the PDG \cite{pdg} listings with
             masses below 1900 MeV/c$^2$.
         \label{fig:scalar_pdg}}
\end{figure}

\noindent
Unquenched lattice QCD computations, summarized in Figure
\ref{fig:scalar_pdg}(a), places the 
lightest Glueball with $J^{PC}=0^{++}$ at about 1.7~GeV.  Unfortunately,
much confusion persists on the existence and nature of scalar states
in or below this region.  They are wide enough to be mixtures of each 
other or with glueball states, so a better understanding of the scalar
meson spectrum would help identify this state.

Fifteen scalar states, shown in Figure \ref{fig:scalar_pdg}(b)
are listed by the Particle Data Group \cite{pdg} (PDG).
The existence of the $\fz(600)$ has long been established, but
its pole position has been difficult to pin down due to interference with
$s$ wave background in $\pi\pi$ and $K\bar{K}$ systems in which it has 
been studied mostly.  There is disagreement on whether this state is 
broad and above, or narrower and below the $\rho$ mass.
The composition of the $J^{PC}=0^{++}$ states $\az(980)$ and $\fz(980)$
- $K\bar{K}$ molecule, $q\bar{q}q\bar{q}$ or $q\bar{q}$ - is unknown.
%
%
%
The spin of the $f_J(1710)$ has been controversial though most
experiments now agree that $J=0$.  Possibly as a result of some
confusion on the identity of the state studied, a few results still 
hint at $J=2$.

Only one strange scalar meson is established to exist, allowing for
only one nonet.  The hint of a further, lower mass $K^{\ast}$ (the
``$\kappa$") seen in the decay of the $D^+$ meson has recently been
published \cite{e791:kpipi}.  Should its existence be confirmed, a
second, light scalar nonet may be established, casting new
light on the identity of the scalar glueball.  This, and other
recent information on the scalar spectrum are the subject of this paper.
%
%
Some prospects for the future are also discussed.

%% file: ddk.tex
Until recently, the majority of knowledge of light quark systems
has come from experiments dedicated to their study.  Many of
these are listed in Table \ref{tab:old_experiments}.

\input{old_experiments.tex}
Recently, decays of $D$ and $D_s^+$
\footnote{Except where indicated otherwise, charge conjugate systems are
implied in this paper.}
mesons to three pseudoscalar
mesons have also begun to provide information.  These decays
often have large branching fractions providing good statistical
accuracy, and generally proceed through intermediate, two meson
systems with natural parity.  Kinematics and angular momentum barrier
factors generally favour scalar ($J^P=0^+$) over vector $1^-$
or tensor $2^+$ systems resulting in an important source of new
information on scalar states.

%% file: old_experiments.tex
\begin{table}[hbt]
 \caption{\sl Experiments that have contributed significantly to
  current knowledge of meson spectroscopy.
 \label{tab:old_experiments}}
 \begin{tabular}{|lll|}
  \hline & & \\[-8pt]
  \bf ``Peripheral model"
         & $\pim\pip\to\pim\pip,~\piz\piz$
         & $\pi N\to$,~E852, \etc.                     \\
  \bf extrapolations:
         & $\pim\pip\to\Km\Kp,~\Kz\Kz$
         & ~~~~~~~~~~"                                  \\
         & $\pim\pip\to\eta\eta,~\eta\eta'$
         & ~~~~~~~~~~"                                  \\
         & $\Km\pip\to\Km\pip$  
         & $\Km p$, LASS                                \\
         & $\Km\Kp\to\Km\Kp,~\Kz\Kz$
         & $\Km p$ (not at pole)                        \\
 & & \\
  \bf $p\bar{p}$ and $Z\bar{N}$
         & $p\bar{p}\to3\piz,~5\piz,~\piz\piz\eta$,
         & Crys. Barrl;
   FNAL \\
  \bf annihilations
         & $~\eta\eta\piz,~\eta\eta'\piz, Z\bar{N}\to$ mesons
         &  E760, Obelix                                \\
 & & \\
  \bf $\phi$ c.f. $\omega$ $s\bar{s}$ content:
         & $J/\psi\to\phi\pi\pi,~\phi K\bar{K},~\omega\pi\pi,~\omega K\bar{K}$
         & Mark II, III                                  \\
 & & \\
  \bf Gluon enriched:
         & $J/\psi\to\gamma\pi\pi,~\gamma K\bar{K},~\gamma\eta\eta,~\gamma\eta\eta'$
         & Mark III                                      \\
         & $pp\to pp+X_{\ssc central}$
         & WA76, WA102                                   \\
         & $\psi'\to J/\psi\pi\pi,~Y'(Y'')\to Y\pi\pi$
         & Mark III                                      \\
 & & \\
  \bf Gluon suppressed:
         & $\gamma\gamma\to\pi\pi,~ K\bar{K}$
         & TPC                                           \\
  \hline
 \end{tabular}
\end{table}

%% file: isobar.tex
In most analyses, $D$ decays to three pseudoscalars $i$, $j$ and $k$
are described as a coherent sum of ``isobar amplitudes"
$\Ampl_{\ssc R}$, each
corresponding to a quasi two body decay \hbox{$D\to R(\to ij)k$}.
$\Ampl_{\ssc R}$ satisfies Lorentz invariance and conserves total spin 
and has the form
$   \Ampl_{\ssc R}(s_{ij}, s_{ik}) =
          F_{\ssc D}(q,r_{\ssc D})
          F_{\ssc R}(p,r_{\ssc R})
          \times BW_{\ssc R}(s_{ij})
          \times~(-2)^J |\vec p|^J |\vec q|^J P_J(\hat p\cdot\hat q)
$
where $\vec p, \vec q$ are momenta of $i$ and $k$ in
the $(ij)$ and $D$ rest frames, respectively.
Form factors $F_{\ssc D}$ ($F_{\ssc R}$) for $D$ ($R$) are parametrized 
in terms of  effective radii $R_{\ssc D}$ ($R_{\ssc R}$) for the decaying
meson and the resonance $R$, respectively.  A Breit Wigner (BW)
propagator
$BW_{\ssc R}~=~\left[s_{\ssc R} - s_{ij} -
               i\sqrt{s_{\ssc R}}\;\Gamma(s_{ij}) \right]^{-1}$
describes the resonance with spin $J$, mass $M_{\ssc R}=\sqrt{s_{\ssc R}}$.
Suffix $R$ denotes a quantity evaluated at $s_{ij}=s_{\ssc R}$.

The distribution of decays in
Dalitz plot coordinates $(s_{ij}, s_{ik})$ (squared invariant mass
combinations) is
  \[ {\cal P}_{\ssc S}(s_{ij}, s_{ik})~=~
     \left|\, a_{\ssc NR}e^{i\delta_{\ssc NR}}~+~\sum_{\ssc R} a_{\ssc R}
     e^{i\delta_{\ssc R}}\Ampl_{\ssc R}(s_{ij}, s_{ik})\, \right|^2 \]
Decays directly to three bodies (NR), not involving a resonance, are
described by a constant, ``contact" amplitude
$a_{\ssc NR}e^{i\delta_{\ssc NR}}$ in the expression above.

Fits are made to obtain values for complex coefficients $ae^{i\delta}$
and resonance parameters.  Experimentally, incoherent backgrounds from
sources other than $D$ decay, and efficiencies affecting the observed
distributions are carefully modelled and incorporated into the fits.

%% file: kappa.tex
\begin{figure}[ht]
 \vskip24pt
 \hbox{\hspace{0.12\textwidth}(a)\hspace{0.39\textwidth}
                             (b)\hspace{0.25\textwidth} (c)}
 \vskip-24pt
 \hbox{%
 \epsfig{file=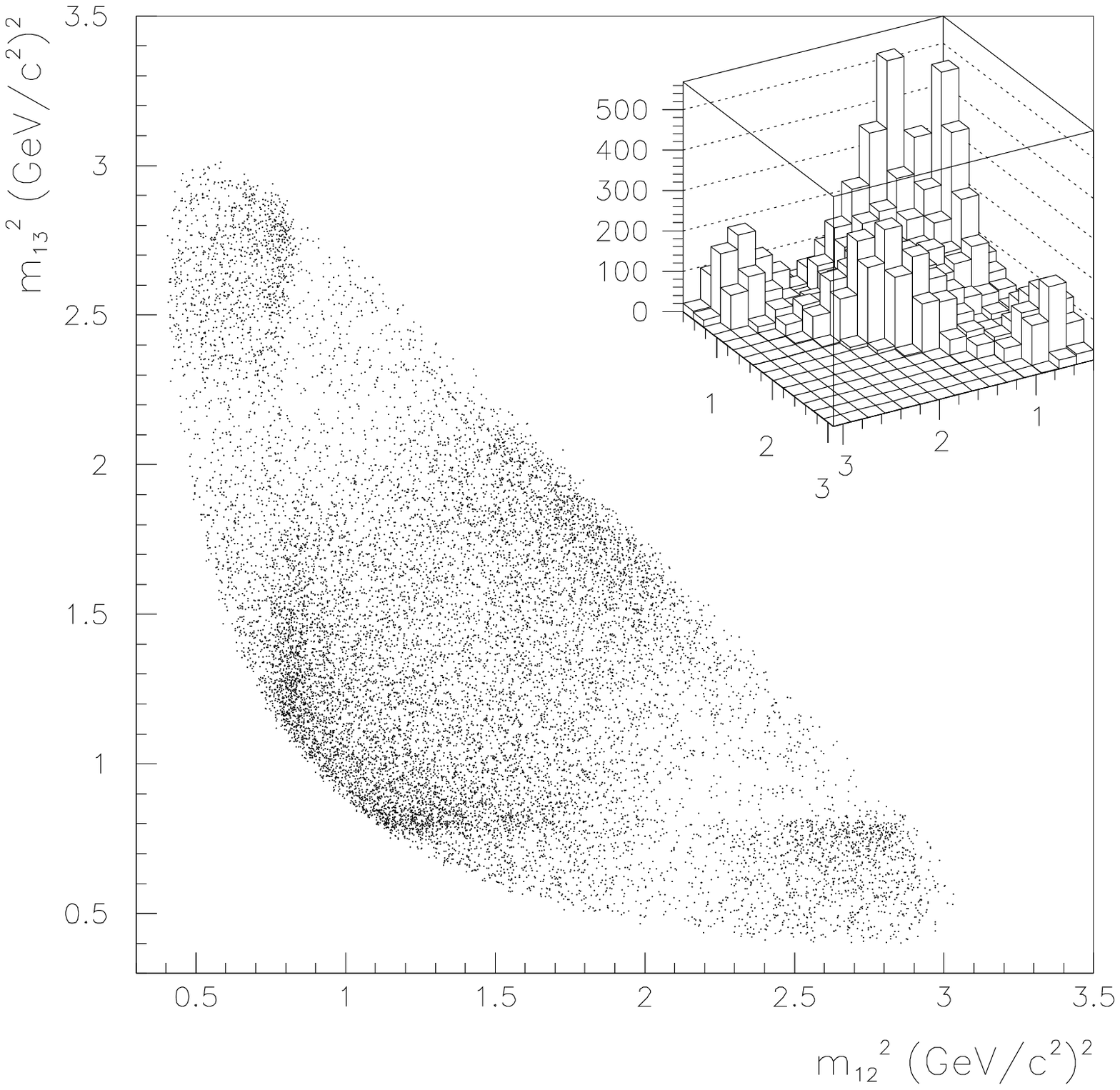,
      width=\setlength{0.38\textwidth},angle=0}
\hskip1.1em
 \epsfig{file=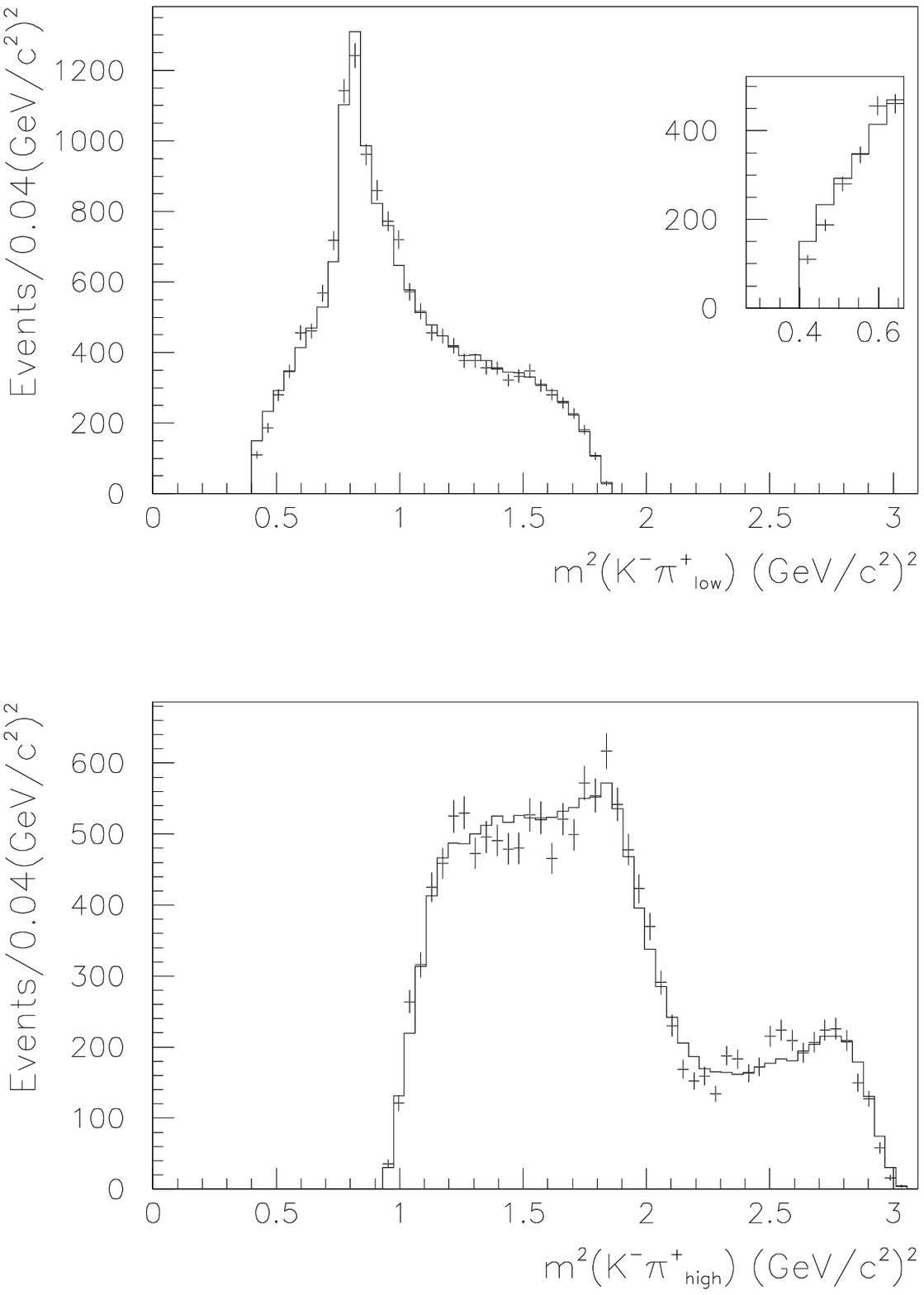,
   width=0.26\textwidth,angle=0}
\hskip1.1em
 \epsfig{file=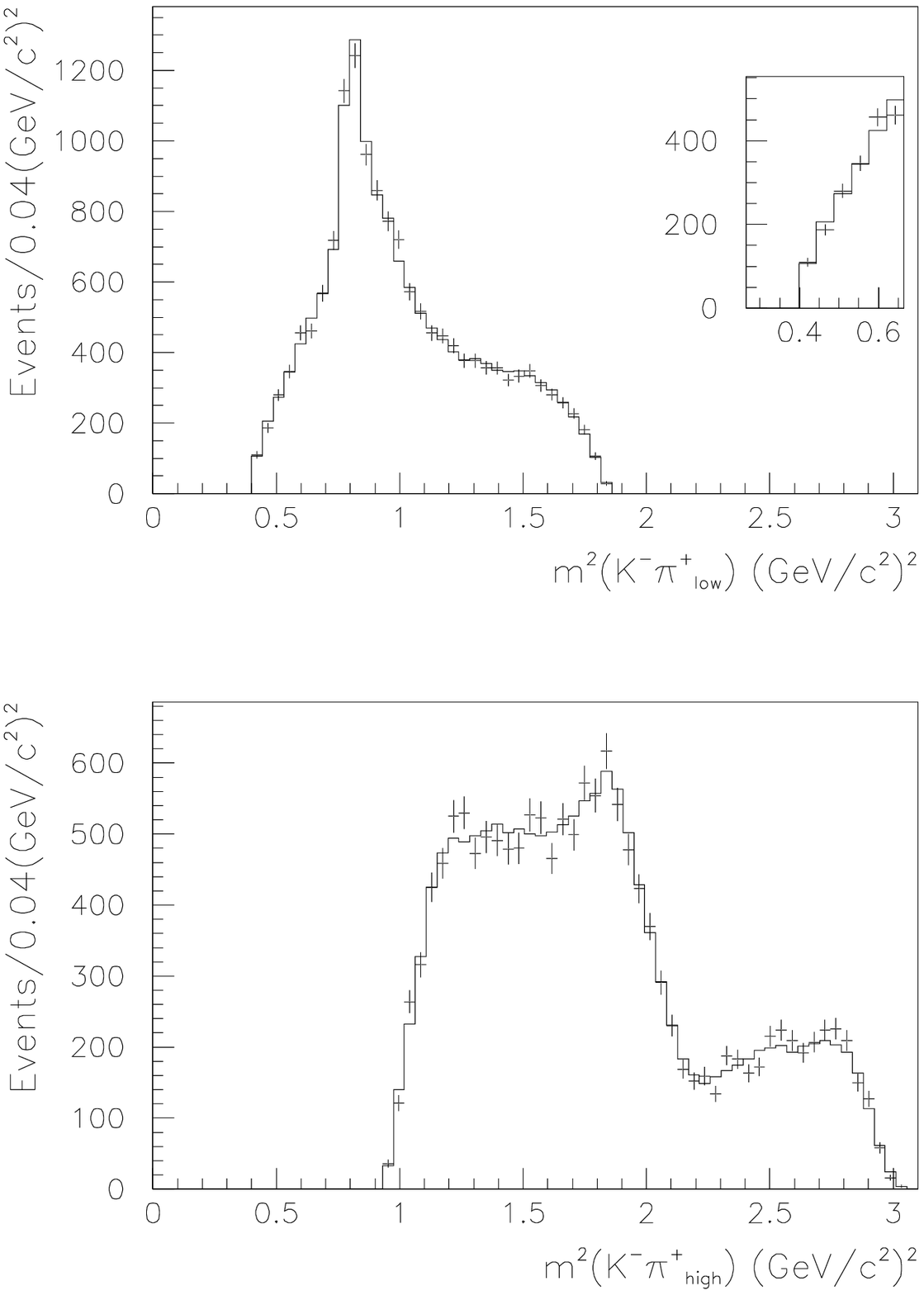,
   width=0.26\textwidth,angle=0}
 }
 \caption{\sl (a) Dalitz plot for 15,090 $D^+\to\Km\pip\pip$
 decays with $\sim$6\% background from E791.
 (b) $\Km\pip$ mass projections showing data (error bars)
 and fit (solid histogram) to model A (no $\kappa$) and
 (c) model B which includes a $\kappa\pi$ amplitude.
 \label{fig:e791_kpipi}}
\end{figure}

E791 reports \cite{e791:kpipi} an isobar model analysis of a
large sample of $D^+\to\Km\pip\pip$ decays.
%
The Dalitz plot in Figure \ref{fig:e791_kpipi}(a)
shows strong $K_1(890)$, $K_{\z}(1430)$, and $K_2(1430)$.
The asymmetry pattern in the $K_1(890)$ indicates
significant, underlying $\Km\pip~s-$wave interference.

Model A, with an NR contribution and only isobars found in the
PDG listings, gives the results in Table \ref{tab:e791_kpipi}.
Masses and widths are fixed at their PDG values.
\begin{table}[h]
  \caption{\sl Fits to the E791 $D^+\to\Km\pip\pip$ Dalitz plot.  Models
  A and B are described in the text.
  \label{tab:e791_kpipi}}
  \[
  \begin{array}{|l|rr|rr|}
  \hline
              & \multicolumn{2}{|c|}{} & \multicolumn{2}{ c|}{} \\[-9pt]
              & \multicolumn{2}{|c|}{\hbox{Model A ($\chi^2$/dof = 167/63)}}
              &
                \multicolumn{2}{ c|}{\hbox{Model B ($\chi^2$/dof = 46/63)}}
                \\
  \hline
              & \multicolumn{1}{c }{} & \multicolumn{1}{c|}{} &
                \multicolumn{1}{c }{} & \multicolumn{1}{c|}{}
  \\[-9pt]
  \hbox{Mode} & \multicolumn{1}{c }{\hbox{\%}}                &
                \multicolumn{1}{c|}{\hbox{phase}}             &
                \multicolumn{1}{c }{\hbox{\%}}                &
                \multicolumn{1}{c|}{\hbox{phase}}                  \\
      \kappa\pip              & \multicolumn{1}{c }{\hbox{-}} &
      \multicolumn{1}{c|}{\hbox{-}}
                              & 47.8\!\pm\! 12.1\!\pm\! 3.7\% &
        187\!\pm\!  8\!\pm\! 17^{\z}                               \\
      \hbox{NR}               & 90.0\!\pm\! 2.6 \% &
        0^{\z}~\hbox{(fixed)}
                              & 13.0\!\pm\!  5.8\!\pm\! 2.6\% &
        349\!\pm\! 14\!\pm\!  8^{\z}                               \\
      K^{\ast}(890)\pip       & 13.8\!\pm\! 0.5 \% &
         54\!\pm\!  2^{\z}
                              & 12.3\!\pm\!  1.0\!\pm\! 0.9\% &
        0^{\z}~\hbox{(fixed)}                                      \\
      K^{\ast}_{\z}(1430)\pip & 30.6\!\pm\! 1.6 \% &
        109\!\pm\!  2^{\z}
                              & 12.5\!\pm\!  1.4\!\pm\! 0.4\% &
         48\!\pm\!  7\!\pm\! 10^{\z}                               \\
      K^{\ast}_{2}(1430)\pip  &  0.4\!\pm\! 0.1 \% &
         33\!\pm\!  8^{\z}
                              &  0.5\!\pm\!  0.1\!\pm\! 0.2\% &
       306\!\pm\!  8\!\pm\!  6^{\z)}                               \\
      K^{\ast}_{1}(1680)\pip  &  3.2\!\pm\! 0.3 \% &
         66\!\pm\!  3^{\z}
                              &  2.5\!\pm\!  0.7\!\pm\! 0.2\% &
         28\!\pm\! 13\!\pm\! 15^{\z)}                              \\
  \hline
    \end{array}
  \]
\end{table}
There are obvious problems with this model, many seen in earlier
analyses \cite{e691:kpipi}.
The NR contribution is large - an effect not usually seen in other
three body $D$ decays - and fractions of the modes sum to
$\sim$140\%.  This indicates significant interference, mostly
with the NR component.  Unlike the earlier analyses, E791's
statistical significance shows that this model gives an unacceptable
fit, with $\chi^2/\hbox{dof} = 2.7$ in 63 bins in the Dalitz plot.
The fit quality is worst in the
low $K\pi$ mass region, and its reflection at $\sim 2.5$~GeV$^2$
as seen in the $\Km\pip$ mass projection in Figure
\ref{fig:e791_kpipi}(b).  In anlyzing $D^{\z}\to\Km\pip\piz$
decay data, the CLEO collaboration \cite{cleo:kmpippiz} also obtain
a poor fit in this region.

Introduction of another scalar resonance, with mass and width
determined by the fit, is required to obtain a good fit.
This fit (Model B) gives the results in Table
\ref{tab:e791_kpipi}, mass projections in Figure
\ref{fig:e791_kpipi}(c) and converges on a mass and width for 
$\kappa$ of $M_{\kappa}=797\pm 19\pm 42~MeV/c^2$ and
$\Gamma_{\kappa}=410\pm 43\pm 85~MeV/c^2$.  The $\kappa$$\pip$
amplitude is dominant and the $NR$ fraction becomes insignificant
($\sim 2\sigma$).  The sum of fractions is $\sim 90$\% and very good
fit quality is found in all Dalitz plot regions.  However, the
$K_{\z}(1430)$ parameters from this fit (Table \ref{tab:k01430}) are
inconsistent with PDG values.
\begin{table}
\caption{\sl New $K^{\ast}_{\z}(1430)$ parameters from E791 and
 LASS collaboration \cite{lass:wmd}
\label{tab:k01430}}
 \[
 \begin{array}{|l|cccc|}
 \hline
      & \hbox{E791 Model B}
      & \hbox{Published}
      & \hbox{LASS Re-fit for}
      & \hbox{PDG}                            \\
      & \hbox{with $\kappa$}
      & \hbox{LASS values}
      & \hbox{$M_{\Km\pip}<M_{K\eta'}$}
      &                                       \\
 \hline
  M_{K_{\z}(\ssc 1430)}
      & 1459\!\pm\! 7\!\pm\! 6
      & 1412\!\pm\! 7
      & 1435\!\pm\! 9
      & 1412\!\pm\!  6~\hbox{MeV/c}^2         \\
  \Gamma_{K_{\z}(\ssc 1430)}
      &  175\!\pm\! 12\!\pm\! 12
      &  294\!\pm\! 40
      &  279\!\pm\! 40
      &  294\!\pm\! 23~\hbox{MeV/c}^2)        \\
\hline
 \end{array}
 \]
\end{table}

\subsection{New $K^{\ast}_{\z}(1430)$ parameters.}

Data for this state are dominated by results from LASS experiment
E135 \cite{lass:k01430} in which it was discovered.  Recently, this
collaboration reports \cite{lass:wmd} that new central values
for mass and width, in Table \ref{tab:k01430}, are obtained when
the fit to their $\Km\pip$ scattering data is limited to the
elastic range (below $K\eta'$ threshold).  This implies that a
larger systematic uncertainty should be attributed to these
parameters than indicated by the PDG listing.

%
%

%% file: sigma.tex

\begin{figure}[hbt]
 \begin{minipage}[ht]{0.30\textwidth}
  \vskip24pt\hskip0.5\textwidth (a)\vskip-24pt
  \centerline{%
  \epsfig{file=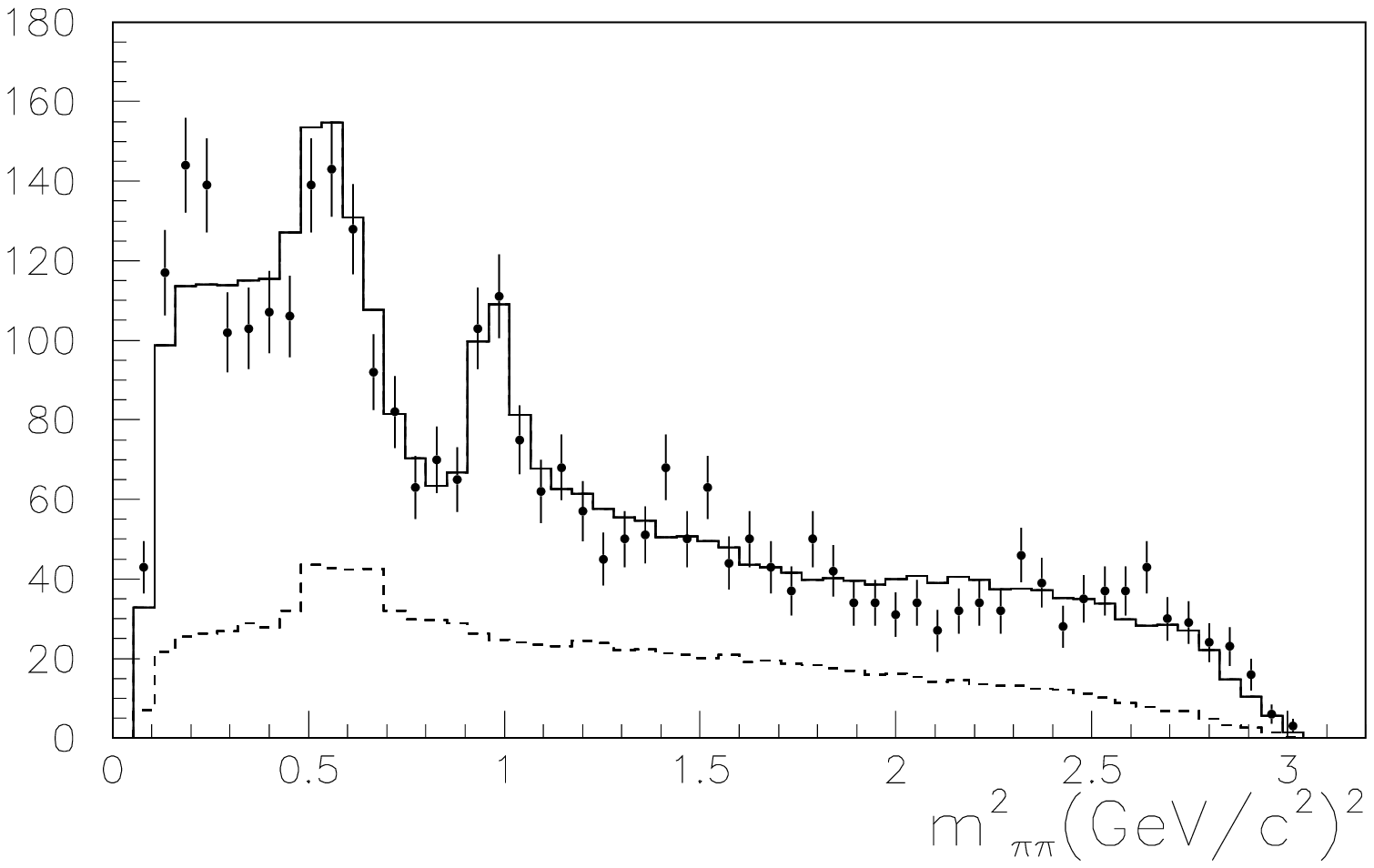,
          height=0.9\textwidth,
          width=\textwidth,
          angle=0}}
  \par\vspace{0pt}
 \end{minipage}
 \hspace{0.0\textwidth}
 \begin{minipage}[ht]{0.30\textwidth}
  \vskip24pt\hskip0.5\textwidth (b)\vskip-24pt
  \centerline{%
  \epsfig{file=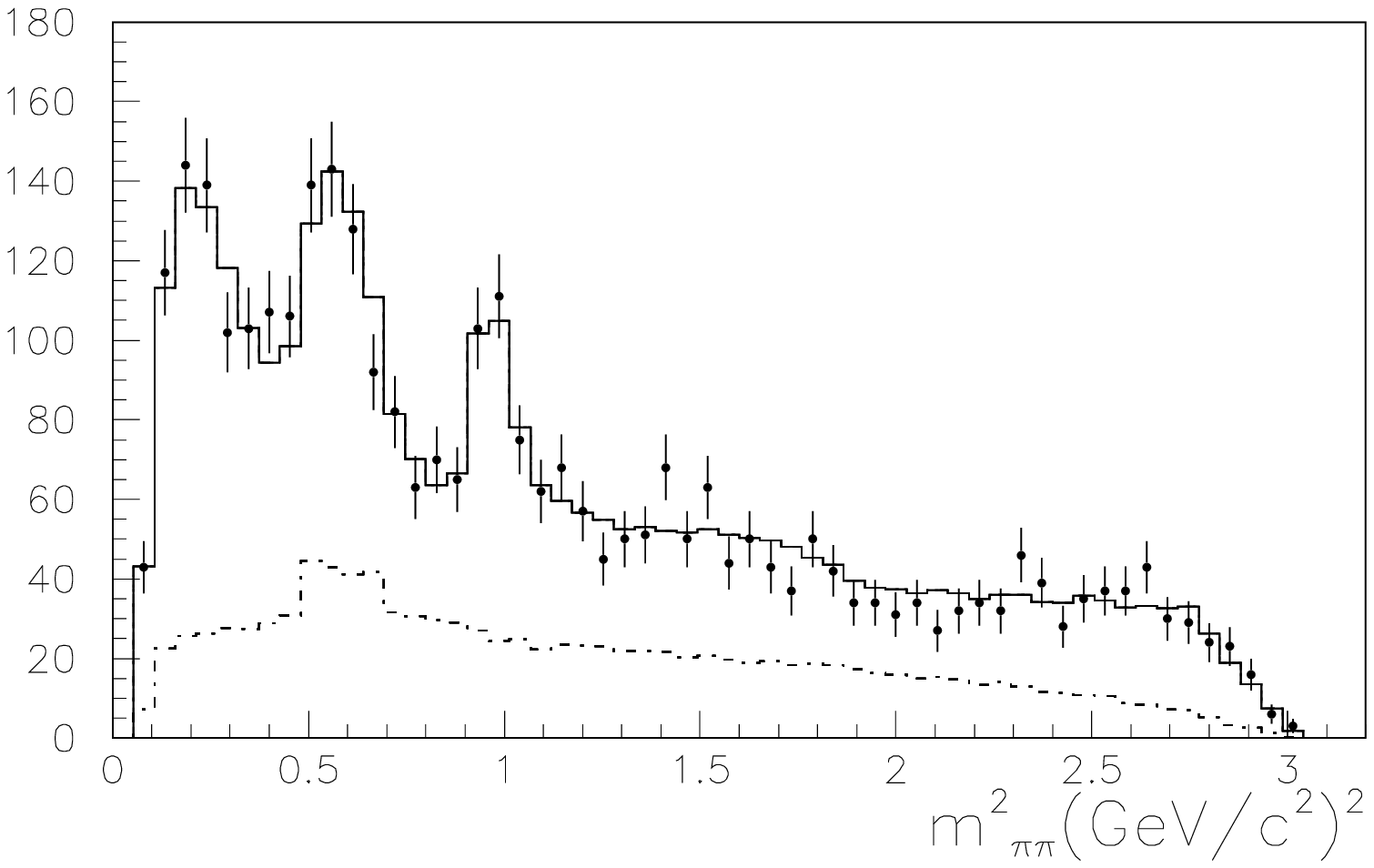,
          height=0.9\textwidth,
          width=\textwidth,
          angle=0}}
  \par\vspace{0pt}
 \end{minipage}
 \hspace{0.0\textwidth}
 \begin{minipage}[ht]{0.30\textwidth}
  \vskip24pt\hskip0.5\textwidth (c)\vskip-24pt
  \centerline{%
  \epsfig{file=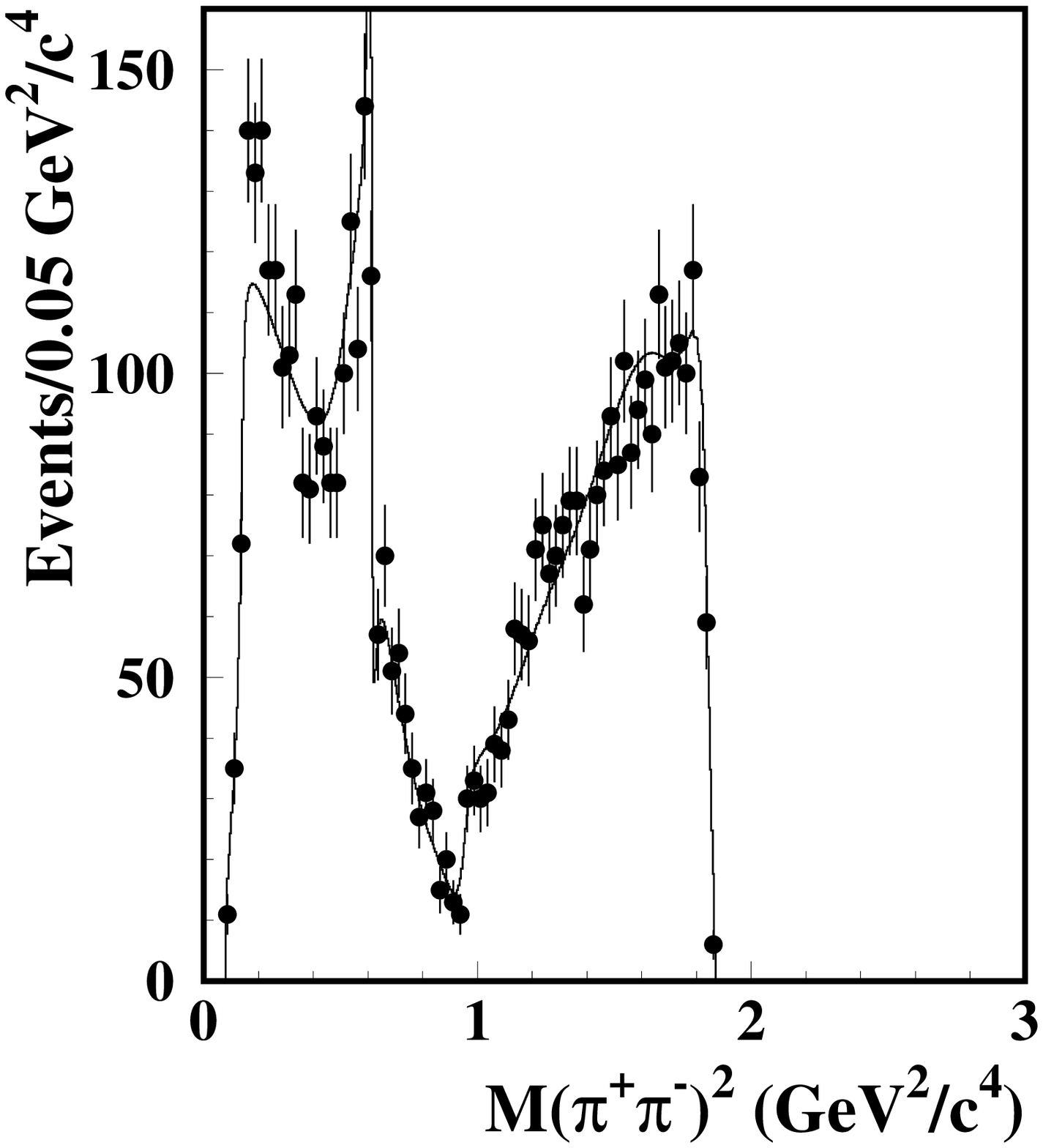,
          height=0.9\textwidth,
          width=\textwidth,
          angle=0}}
  \par\vspace{0pt}
 \end{minipage}
 \caption{\sl (a)  Projection onto the $\pim\pip$ mass axis of
     E791 sample of $1170\pm 65$ $D^+\to\pim\pip\pip$ decays.
     Structure is obvious in the $\rho^{\z}$ and $f_{\z}(980)$
     regions.  There are two entries per event.  Data (error
     bars) and the fit (solid line) to model A described in the
     text are shown.
     (c) $\pim\pip$ mass squared projection from
     $D^{\z}\to\Kz(\bar{K}^{\z})\pim\pip$ decays from ref.
     \cite{cleo:kzpimpip}.  The fit (solid line) shown does not
     include a $\sigma$ amplitude.
 \label{fig:e791_d3pi}}
\end{figure}

Analysis of $D^+\to\pim\pip\pip$ decays by the E791 collaboration
\cite{e791:d3pi} provides an indication of a clear signal that
may correspond to the $f_{\z}(600)$, known as the $\sigma$
meson.  The Dalitz plot was fit to isobars summarized in model A
in Table \ref{tab:e791_d3pi} and shown in the $\pim\pip$ mass
projection in Figure \ref{fig:e791_d3pi}(a).
\begin{table}
 \caption{\sl Results of isobar model fit to $D^+\to\pim\pip\pip$
     Dalitz plot.  Model A and Model B are described in the text.
 \label{tab:e791_d3pi}}
 \begin{center}
 \[
  \begin{array}{|c|rr|rr|}
    \\[-24pt]\hline\\[-12pt]
                        & \multicolumn{2}{|c|}{\hbox{Model A}}
                        & \multicolumn{2}{|c|}{\hbox{Model B}} \\
    \hbox{Mode}         & \hbox{Fraction}
                        & \hbox{Phase}
                        & \hbox{Fraction}
                        & \hbox{Phase}                        \\
    \hline
    & & & &\\[-8pt]
    \sigma\pip          & -
                        & -
                        & 46.3\!\pm\!9.0\!\pm\!2.1\%
                        &  206\!\pm\! 8\!\pm\! 5^{\z}          \\
    \hbox{non resonant} & 38.6\!\pm\! 1.4\%
                        &  150\!\pm\! 12^{\z}
                        &  7.8\!\pm\!6.0\!\pm\!2.7\%
                        &   57\!\pm\!20\!\pm\! 6^{\z}          \\
    \rho(770)\pip       & 20.8\!\pm\! 2.3\%
                        & 0^{\z}~\hbox{(fixed)}
                        & 33.6\!\pm\!3.2\!\pm\!2.2\%
                        &          0^{\z}~\hbox{(fixed)}       \\
    f_{\z}(980)\pip     &  7.4\!\pm\! 4.3\%
                        &  152\!\pm\! 16^{\z}
                        &  6.2\!\pm\!1.3\!\pm\!0.4\%
                        &  165\!\pm\!11\!\pm\! 3^{\z}          \\
    f_{2}(1270)\pip     &  6.3\!\pm\! 3.3\%
                        &  103\!\pm\! 16^{\z}
                        & 19.4\!\pm\!2.5\!\pm\!0.4\%
                        & 57.3\!\pm\!7.5\!\pm\! 5^{\z}         \\
    f_{\z}(1370)\pip    & 10.7\!\pm\! 7.7\%
                        &  143\!\pm\! 10^{\z}
                        &  2.3\!\pm\!1.5\!\pm\!0.8\%
                        & 105.4\!\pm\!18\!\pm\!0.6^{\z}        \\
    \rho(1450)\pip      & 22.6\!\pm\! 2.1\%
                        &   46\!\pm\! 15^{\z}
                        &  0.7\!\pm\!0.7\!\pm\!0.3\%
                        &  319\!\pm\!39\!\pm\!11^{\z}          \\
   \hline
  \end{array}
\]
\end{center}
\end{table}
In this model, no $\sigma\pip$ amplitude is included.  The fit
is poor with $\chi^2\sim 80$ for 63 degrees of freedom.  The NR
decay is dominant and the amplitudes for $\rho(1450)$ and $\rho(770)$
are almost equally strong - an odd situation.  The fit is
particularly bad in the low mass $\pip\pim$ region.  These results
are generally compatible with the only previous measurements of
this decay mode by the E687 collaboration \cite{e687:d3pi}.

In model B, a $\sigma\pip$ amplitude with scalar BW parameters,
$M_{\sigma},~\Gamma_{\sigma}$, allowed to float freely, is
introduced.  Values for $M_{\sigma},~\Gamma_{\sigma}$ that
result are in Table \ref{tab:sigmas}, indicating a $\sigma$
below the $\rho(770)$.
%
\begin{table}[hbt]
 \caption{\sl Fits to various neutral dipion systems.
 \label{tab:sigmas}}
 \begin{center}
 \begin{tabular}{|l|lccc|}
 \multicolumn{5}{c}{}\\[-36pt]
 \hline\\[-12pt]
 \bf Channels
    & \bf Data
    & $\mb M_{\sigma}$~MeV/c$^2$
    & $\mb\Gamma_{\sigma}$~MeV/c$^2$ 
    & \bf Low Mass \\
 \hline & & & & \\[-12pt]
 $D^+\to\pim\pip\pip$$^{\mb\dagger}$
    & E791
    & $478^{+24}_{-23}\pm 17$
    & $324^{+42}_{-40}\pm 21$
    & enhanced                         \\
 $\tau\to\nu_{\tau}\pim
      \left(\pi^{-\z}\pi^{+\z}\right)$
    & CLEO
    & $860^{\dagger}$
    & $880^{\dagger}$
    & enhanced                         \\

 $D^{\z}\to\Kz(\bar{K}^{\z})\pim\pip$
    & CLEO
    & $478^{\mb\ast}$
    & $324^{\mb\ast}$
    & enhanced                         \\
 $\phi\to\piz\piz\gamma$
    & KLOE
    & $478^{\mb\ast}$
    & $324^{\mb\ast}$
    & enhanced                         \\
 $J/\psi\to\omega\pi^{\z-}\pi^{\z+}$
    & DM2
    & $482\pm  3$
    & $710\pm 30$
    & enhanced                         \\
 $pp\to\piz\piz$ (central)
    & GAMS
    & $590\pm 10$
    & $325\pm 10$
    & enhanced                         \\
 \hline & & & & \\[-12pt]
 $J/\psi\to\phi\pi\pi,~\phi K\bar{K}$ & Mark II
    & & & suppressed                     \\\vspace{2mm}
 $\Upsilon(2S)\to\Upsilon(1S)\pi\pi$,
 & \parbox[h]{2.5cm}{\small\vspace{1mm}CLEO, ARG,\\
                      \vspace{-3mm}CUSB, Cr Ball\vspace{2mm}}
    & & & suppressed                     \\
 $\Upsilon(3S)\to\Upsilon(2S)\pi\pi$, & CLEO
    & $526^{+ 48}_{- 37}$
    & $301^{+145}_{-100}$
    & suppressed                     \\
 $\Upsilon(3S)\to\Upsilon(1S)\pi\pi$, & CLEO
    & & & enhanced                         \\
 $\psi(2S)\to J/\psi\pi\pi$           & Cr Ball
    & & & suppressed                     \\
 \hline & & & & \\[-12pt]
 $I=0~s~\hbox{wave}~\pi\pi\to\pi\pi$  &
    & $602\pm 26$
    & $392\pm 54$
    & suppressed                     \\
 \hline
 \multicolumn{5}{l}{%
 $^{\dagger}$ Values fixed at prediction of Tornqvist,
 Z. Phys. C68, 647 (1995).} \\
 \multicolumn{5}{l}{%
 $^{\ast}$ Values from E791 used in the fits.}
 \end{tabular}
 \end{center}
\end{table}
The fit, whose results are in Table \ref{tab:e791_d3pi}, is of
a significantly improved quality, $\chi^2\sim 57$ for 63
degrees of freedom.  It describes the low mass $\pim\pip$ mass
region shown in Figure \ref{fig:e791_d3pi}(b) well.
The $\sigma\pip$ mode dominates the decay - but the $NR$ amplitude
becomes neglibibly small as does that for the $\rho(1450)\pi$.

\section{Comments on E791 $\sigma$ and $\kappa$ Signals.}

Tests made by E791 reveal that the scalar BW phase motion is
important in obtaining acceptable fits \cite{e791:d3pi, e791:kpipi}.
Fits to a ``real BW" (a peak, no phase motion) result in poor 
$\chi^2$ and large sum of resonant fractions.  Fits to vector 
and tensor forms are also significantly worse.
Nevertheless, the isobar model used by E791 - with a scalar BW for
the $\sigma$ and $\kappa$ - may not be formally correct.
\footnote{Unitarity is ignored.  It may be necessary to include
constraints from $\pi\pi$ or $\Km\pip$ elastic scattering and
from chiral symmetry.}

E791 plans a {\sl model independent measurement} of the
$s$ wave $\pim\pip$ and $\Km\pip$ magnitudes and phases as a
function of mass, using interference between the two
identical $\pip$ (Bose symmetrized) amplitudes in these $D$
decays.  This could help resolve whether or not poles really
exist in these systems.

What can certainly be said is that E791 data clearly
indicate the need for some phase motion in the $s$ wave meson-
meson systems and that a scalar BW is one model that works well.
The possibility that another parametrization, possibly with
no scalar states at all, could also fit the data satisfactorily
is not excluded.

\section{Other evidence for $\sigma$.}

\input{other_sigma.tex}

%% file: other_sigma.tex
In $D^{\z}\to\Kz(\bar{K}^{\z})\pim\pip$ decays, CLEO
\cite{cleo:kzpimpip}
note that an acceptable fit requires a $\Kz\sigma$
contribution with mass and width similar to those found in E791.
A fit without $\sigma$, shown in figure \ref{fig:e791_d3pi}(c),
fails to account for the low $\pim\pip$ mass region.

Other instances where a $\sigma$ pole is added in the
description of the neutral di-pion
systems are summarized in table
\ref{tab:sigmas}.
The CLEO collaboration \cite{cleo:tau} observe, in
 $\tau\to\nu_{\tau}\pim\left(\pi^{-\z}\pi^{+\z}\right)$ decays,
that the
three pion systems (dominantly $J^P=1^+$) require a $\sigma\pim$
amplitude to obtain an acceptable fit.
%
In analyzing $\phi\to\piz\piz\gamma$ radiative decays, the
KLOE collaboration found the best fit among those tried
was obtained if a
$\sigma$ with parameters taken from E791 is included.

Low mass enhancements are observed in di-pion systems in
$J/\psi\to\omega\pi^{\z-}\pi^{\z+}$,
$\Upsilon(3S)\to\Upsilon(1S)\pim\pip$ decays and $\piz\piz$
from $pp$ central production.  However, in
other cases, the
low mass di-pion system is suppressed.
The ``sigma collaboration"
\footnote{M. Y. Ishida, S. Ishida, T. Ishida,
 T. Komada, A. M. Ma, H. Shimizu,
 K. Takamatsu, T. Tsuru \\
 Tokyo Inst. Tech., Nihon U, KEK, IHEP Beijing,
 Yamagata U., CROSS.
 They also fit \cite{sigma:elastic} $s$ wave $\pi\pi$ and
 $\Km\pip$ elastic scattering data to this model with a
 ``background"
 with a falling phase to accomodate the $\sigma$ and $\kappa$.
}
suggest that all these data can be fit with a model where
interference between contact and $\sigma$ pole terms can cause
either enhancement or suppression
\cite{sigma:tsuru, sigma:upsilon}.
Their fits to
data in these channels, in table \ref{tab:sigmas} indicate that
in these systems, $\sigma$ masses group around the
mass region $\sim 500-600$~MeV/c$^2$.  Results from BES, with
58M $J/\psi$ should be an interesting test for these ideas
which are not universally accepted in the theoretical community.

%% file: f0.tex
More information on the $f_{\z}(980)$ and $a_{\z}(980)$ have
recently come from measurements at KLOE, SND and CMD-2 of
radiative decays of $\phi$.  These results were reviewed in
this conference \cite{antonelli}.  The radiative transition
branching fractions are about an order of magnitude larger than
expected for pure $s\bar{s}$ or $K\bar{K}$ composition,
possibly indicating significant $q\bar{q}q\bar{q}$ content.

\begin{figure}[hbt]
\vbox{%
\begin{minipage}[ht]{0.48\textwidth}
 \vskip24pt
 \hbox{\hspace{0.27\textwidth}\bf (a)}
 \hbox{\hspace{0.26\textwidth}\bf E791}
 \vskip-36pt
 \centerline{%
 \epsfig{file=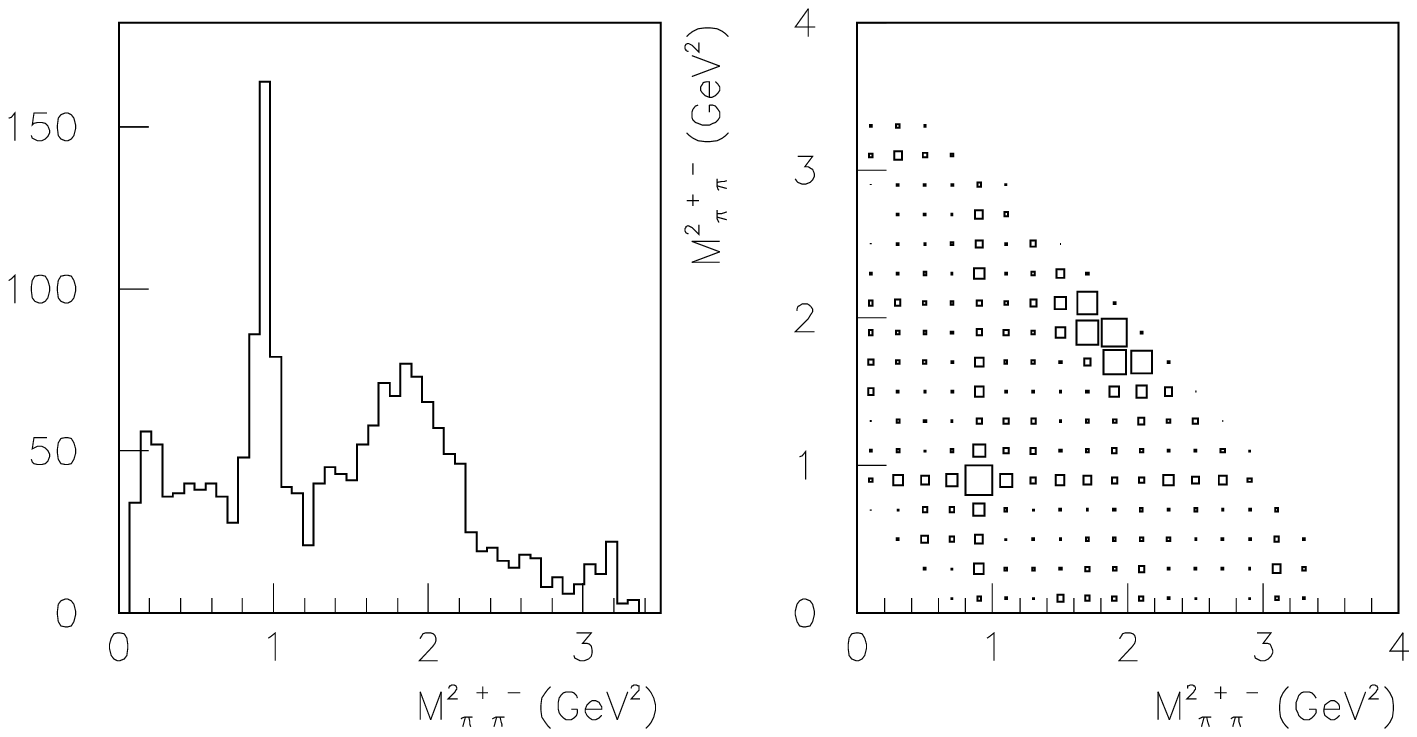,
      width=0.98\textwidth,angle=0}}
\par\vspace{0pt}
\end{minipage}
\hskip0.05\textwidth
\begin{minipage}[ht]{0.47\textwidth}
 \vskip24pt
 \hbox{\hspace{0.22\textwidth}\bf (b)}
 \hbox{\hspace{0.20\textwidth}\bf FOCUS}
 \vskip-36pt
 \epsfig{file=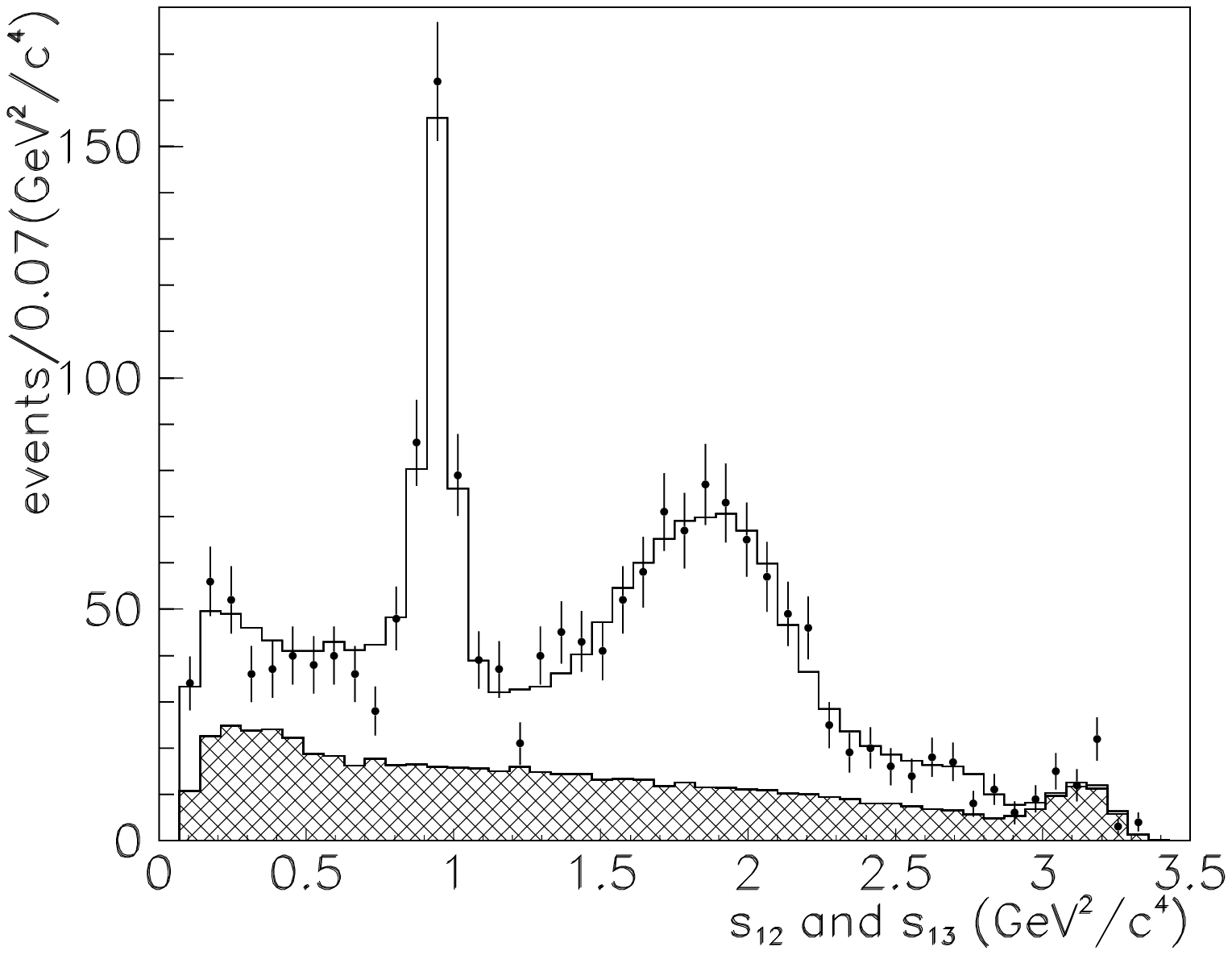,
      height=0.47\textwidth,
      width=0.46\textwidth,angle=0}
 \hskip0.05\textwidth
 \vskip-0.54\textwidth
 \hskip0.49\textwidth
 \epsfig{file=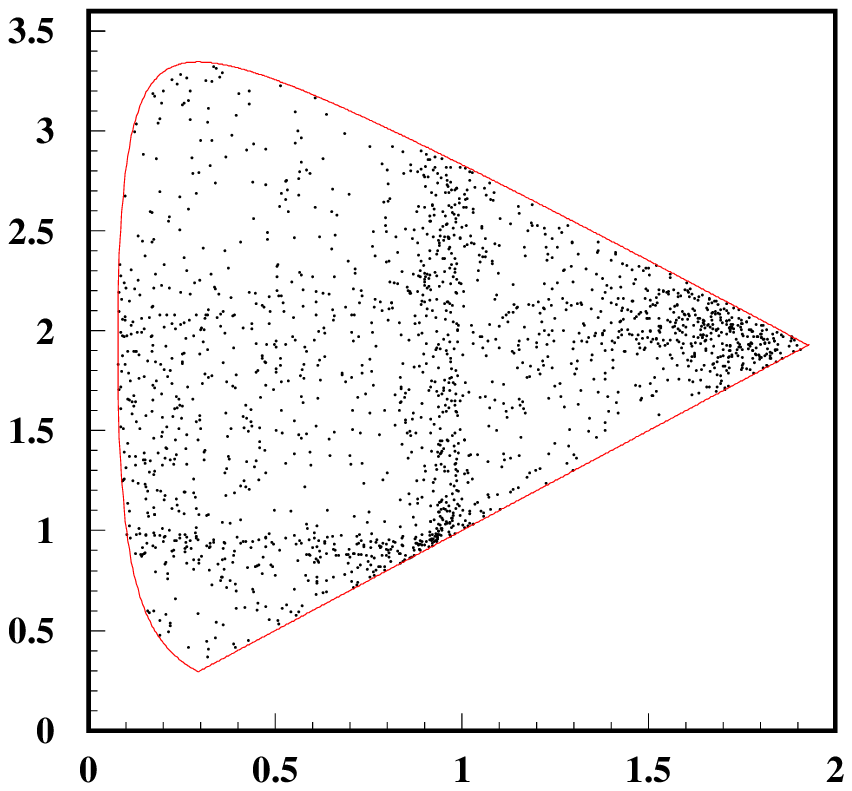,
      width=0.47\textwidth,
      height=0.30\textwidth,angle=-90.0}
 \\
%
\par\vspace{0pt}
\end{minipage}
\vskip0pt
%
\begin{minipage}[ht]{0.45\textwidth}
 \vskip24pt
 \hbox{\hspace{0.10\textwidth}\bf (c)
       \hspace{0.57\textwidth}\bf BaBar}
 \vskip-24pt
 \centerline{%
 \epsfig{file=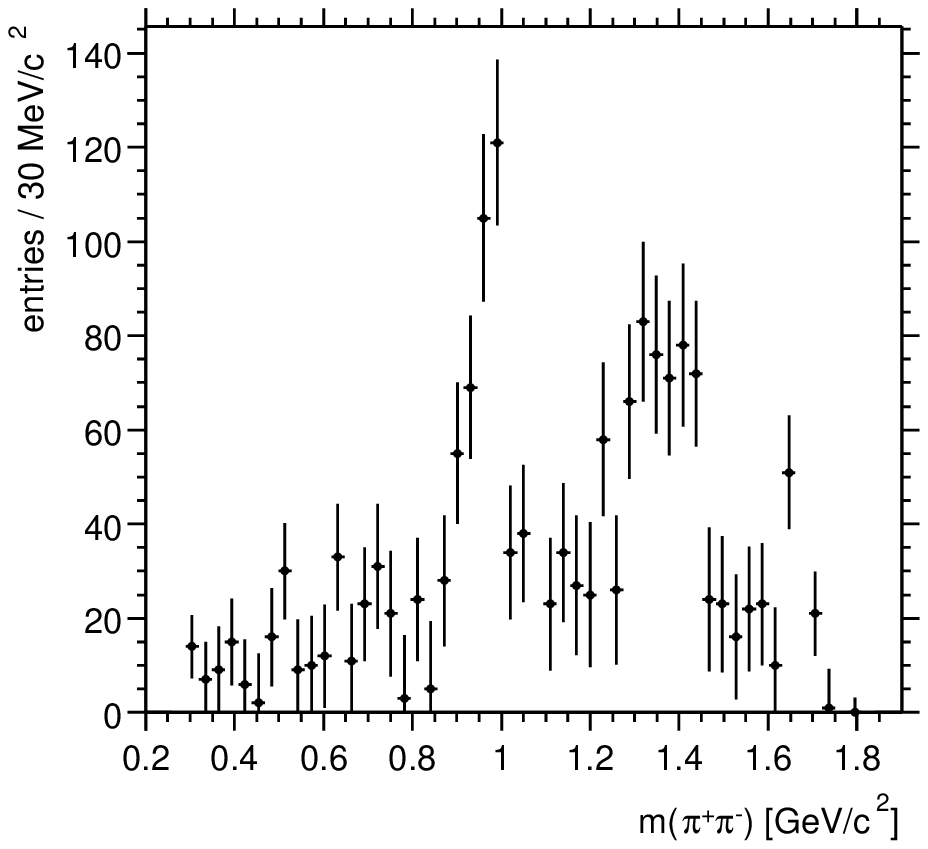,
      height=\setlength{0.49\textwidth},
      width=0.49\textwidth,angle=0}
\hskip0.02\textwidth
 \epsfig{file=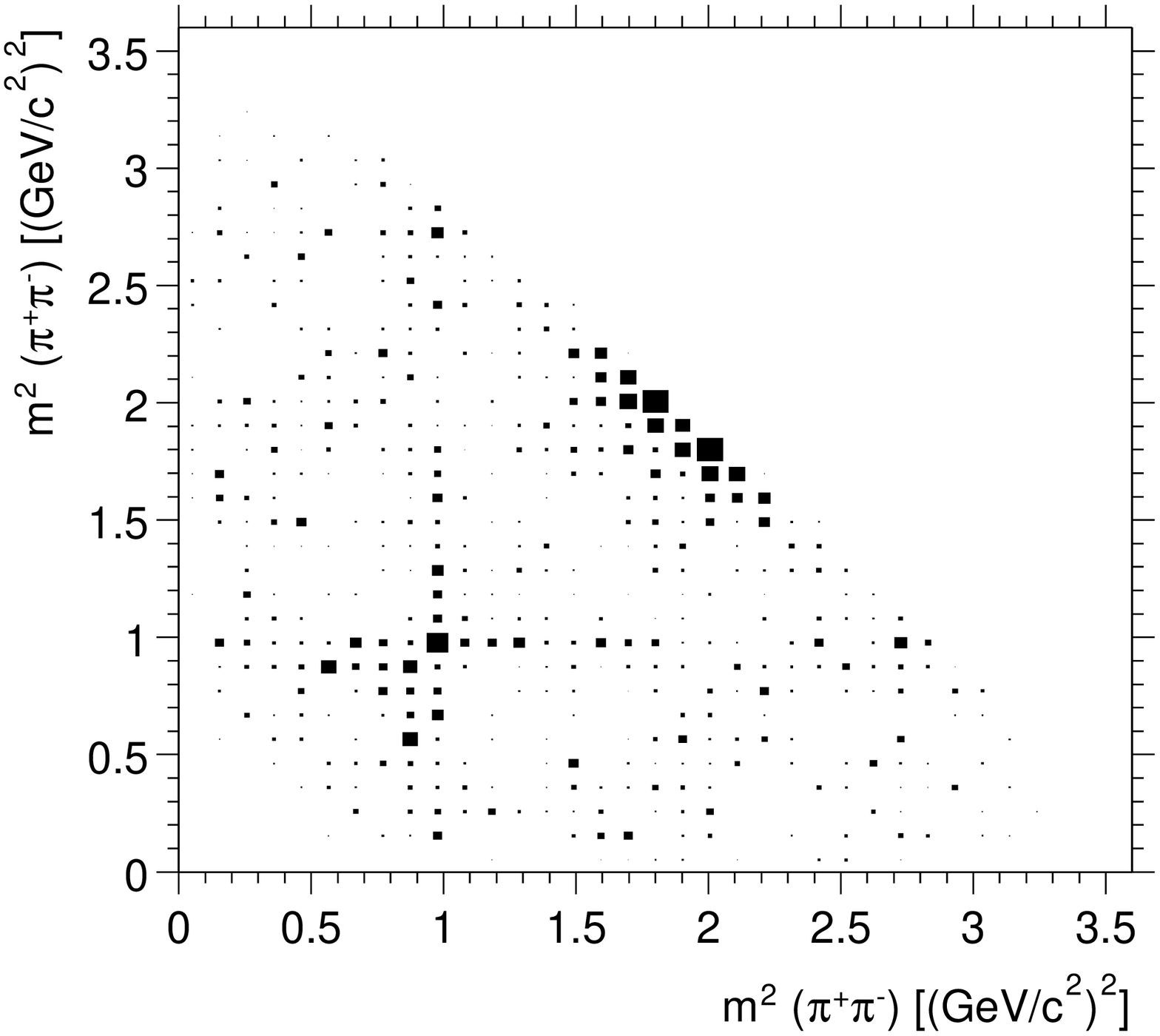,
      height=\setlength{0.49\textwidth},
      width=0.49\textwidth,angle=0}
 }
\par\vspace{0pt}
\end{minipage}
\hskip0.05\textwidth
}
\caption{\sl $\pim\pip$ effective mass distributions and
    Dalitz plots from $D_s^+\to\pim\pip\pip$ decays for
    (a) E791 ($848\pm 44$ events);
    (b) FOCUS ($1445\pm 50$ events); and
    (c) BaBar ($\sim$800 events).
    \label{fig:dsplots}}
\end{figure}

$D_s^+\to f_{\z}(980)\pi^+$ decays which would be expected
to reveal information on the $s\bar{s}$ component of
$f_{\z}(980)$ have also been examined by both the E791 \cite{e791:ds3pi}
and FOCUS \cite{focus:ds3pi} collaborations.  BaBar also plans to
use their large data sample for this.
In Figure \ref{fig:dsplots}(c), $\pim\pip$ mass spectra and
Dalitz plots for $D_s^+\to\pim\pip\pip$ events are shown.
Unlike the E791 and FOCUS plots, the BaBar data are
a preliminary sample ($\sim 20 fb^{-1}$) and no results are yet
available.
The $f_{\z}(980)$ signals are seen as clear peaks on a small
background, in contrast with observations in $\pi\pi$ and $KK$
scattering where, due to the underlying background phase,
the state usually appears as a dip in the cross section.
\setlength{\fh}{\textwidth}
%


Both $\fz(980)$ and $\az(980)$ have a line shape complicated by
proximity to $K\bar{K}$ threshold.  It is described approximately
by
$  CC(s)  = 
    \left[s-m_{\z}^2+im_{\z}\left(\Gamma_K+\Gamma_{\pi}\right)\right]^{-1}
$
where
$    \Gamma_{\pi}  =  g_{\pi}\sqrt{s/4-m_{\pi}^2}~~~
    \Gamma_K  =  {g_K\over 2}\left(\sqrt{s/4-m_K^2}+\sqrt{s-m_{\Kz}^2}\right)
$
In the fits to E791 and FOCUS data, account was taken of this, and
measurement of the ratio of the $\pi\pi$ and $K\bar{K}$ couplings $g_K$
and $g_{\pi}$ was attempted.  In E791 this line shape was fitted directly
and in FOCUS a K matrix fit was used.

\begin{table}
 \caption{\sl The $\fz(980)$ parameters from the statistically most
       significant experiments.  Systematic uncertainties
       are included, where given, in parentheses.  Experiments are
       labelled as $D_s$ decay (A), $pp$ central production (B) or
       $\phi$ radiative decay (C).
 \label{tab:f0980}}
 \begin{center}
 \begin{tabular}[h]{|l|cccccc|}
 \multicolumn{7}{c}{} \\[-12pt]
 \hline & & & & & & \\[-12pt]
       & $M_{\z}$ 
       & $\Gamma_{\z}$
       & $g_K$
       & $g_{\pi}$
       & $g_K/g_{\pi}$
       &                             \\
 \hline & & & & & & \\[-12pt]
 E791  & $977\!\pm\!3 (2)$
       & $44\!\pm\!2 (2)$
       & $0.02 \!\pm\! .04 (.03)$
       & $0.09 \!\pm\! .01 (.01)$
       & $0.22\!\pm\!.44$
       & A                           \\
 FOCUS & $982\!\pm\!30$
       & $89$ to $ 32$
       & -
       & -
       & $2.09\!\pm\!.53$
       & A                           \\
 WA76  & $979\!\pm\!4$
       & $72\!\pm\!8$
       & $0.56 \!\pm\! 0.18$
       & $0.28 \!\pm\! 0.04$
       & $2.00\!\pm\!.70$
       & B                           \\
 WA102 & $987\!\pm\!6 (6)$
       & $48\!\pm\!12 (8)$
       & $0.19 \!\pm\! .03 (.04)$
       & $0.40 \!\pm\! .04 (.04)$
       & $2.10\!\pm\!.62$
       & B                           \\
 KLOE  & $973\!\pm\!1$
       & -
       & $2.79\!\pm\!0.12$
       & -
       & $4.00\!\pm\!.14$
       & C                            \\
 CMD2  & $975\!\pm\!7 (2)$
       & -
       & $1.48\!\pm\!0.32$
       & -
       & $3.61\!\pm\!.62$
       & C                            \\
 SND   & $969\!\pm\!5$
       & -
       & $2.47\!\pm\!0.73$
       & -
       & $4.40\!\pm\!.8$
       & C                             \\
 \hline
 \end{tabular}
 \end{center}
\end{table}
Results are summarized in Table \ref{tab:f0980} where they can be compared
with $pp$ central production and radiative $\phi$ decay results.
Mass and width parameters from a simple $s$ wave BW are also given.
There is considerable disagreement in these parameters, even between
E791 and FOCUS.  Apparently, large systematic effects arise both from
the various production mechanisms and the fit methods.
These differences may be due in part to assumptions made in background
$\pi\pi$ $s$ wave shapes in $pp$ central production and $\phi$ radiative
decays.  Probably the difficulty in including effective mass resolution
in line shapes in the $D_s$ fits also plays a role
\footnote{%
Perhaps striking is that the signals observed in $D_s$ data, whose
spectator model decays would be expected to produce an $s\bar{s}$
system, is narrow.  If this signal were an $s\bar{s}$
state, the preferred decay to $K\bar{K}$ would be kinematically
restricted, making the state narrow.  One is tempted to question whether
or not the $f_{\z}(980)$ is really a unique state.}.

What is needed in future $D_s$ meson studies with larger samples
is a coupled channel approach including $\pi\pi$, $K\bar{K}$ and 
$\eta\pi$ decay modes and the $a_{\z}(980)$.  The BaBar collaboration 
plans such an approach and this will hopefully help in sorting out 
this confusing situation.

%% file: f0_other.tex
Both FOCUS and E791 see evidence for an additional $f_{\z}$
signal at a mass above the $f_{\z}(980)$.  In fitting their Dalitz
plots, mass and width parameters for a scalar BW for this isobar
were allowed to float.  Other discrepancies exist between the $D_s$
results from the two experiments, but they agree quite well on mass and
width of this $f_{\z}$.
\begin{table}
  \caption{\sl ``$f_{\z}(1370)$" parameters from fits to
      $D_s^+\to\pim\pip\pip$ decays from
      E791 and FOCUS experiments.
  \label{tab:f01450}}
  \begin{minipage}[ht]{\textwidth}
  \begin{center}
  \begin{tabular}{|l|ccc|}
   \hline & & & \\[-12pt]
                      &\bf   E791
                      &\bf   FOCUS
                      &\bf   PDG            \\
   \hline & & & \\[-12pt]
   $M_{\z}$~(MeV/c$^2$)
                      & $1434 \pm 18 \pm 9$
                      & $1473 \pm  8$
                      & 1200-1500        \\
   $\Gamma_{\z}$~(MeV/c$^2$)
                      & $172 \pm 32 \pm 6$
                      & $112 \pm 17$
                      & 200-500          \\
   \hline
  \end{tabular}
  \end{center}
  \end{minipage}
\end{table}
It is not clear this state can be identified with $\fz(1370)$.
Measurements in this mass range from $\pi\pi\to\pi\pi$,~$K\bar{K},~
\eta\eta,~\sigma\sigma$, \etc scattering have suffered from
interference with a large, uncertain $s$ wave background and
indicate a broad pole near 1370 MeV/c$^2$ whose parameters
depend on interference with the narrower $f_{\z}(1500)$.  Neither
E791 nor FOCUS find much evidence for $\fz(1500)$ in the $D_s$ fits.

The clean $f_{\z}(980)$ signal observed in these decays suggests
that a clearer interpretation of pole positions of $f_{\z}$ states
may be possible than before.  However, this seems far from
realization at this stage.

%% file: summary.tex
The hint of a $\kappa$ state in E791 is an important
development.  Equally important are a growing number of
instances where a low mass, relatively narrow $\sigma$
amplitude can describe data that comes from a number of
sources not examined in this way before.  A number of
discrepancies in $f_{\z}$ parameters do remain, however.

These observations have required large samples of data.
Hopefully they will be better understood when even more
data, in other channels and in other charge states, are
analyzed.  These should come from FOCUS, BaBar and
BELLE, BES, GSI and CLEO C in the foreseeable future.

Hopes for progress in defining the scalar spectrum
hinge on the proof that a $\kappa$ pole really exists and
on finding a reliable way to determine both $\sigma$ and
$\kappa$ pole parameters.  More data may come, but
a consensus on the
way to describe these observations and also $s$ wave
$I=1/2~K\pi$ and $I=0~\pi\pi$ scattering data in a
consistent way is badly needed.
